\documentclass[a4paper]{article}
\usepackage{graphicx}
\usepackage{amsmath}

\begin{document}

\begin{center}
{\bf \Large
Transition from Uniform Motion to Stick-Slip in the Rice-Ruina Model of One and Two Blocks
}\\[5mm]

{\large
Jacek Szkutnik and Krzysztof Ku{\l}akowski
}\\[3mm]

{\em
Department of Applied Computer Science,
Faculty of Physics and Applied Computer Science,
AGH University of Science and Technology\\
al. Mickiewicza 30, PL-30059 Krak\'ow, Poland
}

\bigskip
\today
\end{center}

\begin{abstract}
We analyze the Rice-Ruina state and rate dependent friction model. The system consists of 
one or two blocks driven by springs with constant velocity on a dry, rough surface. Our discussion 
is limited to the creep-like motion, when the driving velocity is small. Two regimes of motion are
observed: stick-slip and steady sliding. The stability of the steady sliding depends on 
the model parameters. Numerical and analytical results show a transition between two regimes: 
the system passes directly from uniform to stick-slip motion. The calculations are performed 
also for two driven blocks. Then, the transition has the same character and it appears at the same 
point.
\end{abstract}



\section{Introduction}

Friction between solids is a complex problem, first investigated in 1699 by Amontons \cite{bur}.
In 1781, Coulomb formulated his laws, known as Amontons-Coulomb laws. Suppose that a solid block
pressed to a flat surface by a normal force $W$, and the contact surface is of nominal area $S$. 
The static and dynamic friction coefficients $\mu _s$ and $\mu _d$ as the proportionality 
coefficients between the friction forces $F_s$ and $F_d$ and the force $W$. $F_s$ is the force 
necessary to move a standing block. $F_d$ is the friction force during a uniform motion. The laws
state that both friction coefficients do not depend on $W$ and $S$, and that $\mu_d<\mu_s$.
These laws are widely used until now \cite{bur}.

In recent years, interest on friction has been revived because of its possible relevance for
earthquakes \cite{car}, and of its obvious relevance for technology \cite{book}. Detailed 
experimental studies revealed subtle effects on nanoscopic scale, and several phenomenological
models have been formulated \cite{bt,rabi,rr,hes}. In particular, the creep regime is of interest,
which can be considered as an intermediate stage between motion and rest. The idea is that during 
a very slow motion, the interface between solids preserves to some extent the information
on history of contact points. For this regime, the model of Rice and Ruina \cite{rr} is considered
to be appropriate \cite{bau,bauII}. Some variation of this model was formulated in \cite{hes}.

In these models, the physical system is a block connected to a driving mechanism by a spring. 
The mechanism drives 
the block with a constant velocity. The uniform motion of the block is not a unique solution. An
alternative is the so-called stick-slip motion, when the block velocity is varying periodically 
from zero to some maximum, then back to zero and so on. There is an experimental evidence that
once the uniform phase loses its stability, the motion becomes oscillatory, and the amplitude
of the oscillations increases gradually from zero when the control parameter departures the 
transition \cite{bur,hes,bauII,ssc}. This suggests an appearance of the supercritical Hopf 
bifurcation \cite{glen}. However, as it was pointed out in \cite{bau2}, further corrections to the 
Rice-Ruina equations are necessary to reproduce the gradual increase of the amplitude of the
oscillations, present in the Hopf bifurcation. Without these corrections, the transition
leads directly from the uniform motion to the stick-slip effect. In other words, the Rice-Ruina
equations are not generic; a slight modification of the model leads to finite oscillations,
and the Hopf bifurcation is observed. Examples of such modifications are described in 
\cite{bau2,lim}; more general discussion of possible memory effects in the problem of friction 
can be found in \cite{ran}. However, the Rice-Ruina equations are in most cases sufficient
for a convenient description of the uniform motion in the creep regime. 

In realistic situations, the number of contact points between the moving surfaces is larger than 
one, and it depends on the normal force \cite{book,rabi,bt}. This is an indication of a gap between 
theory and experiment on friction. It is clear that a more appropriate and general model should 
contain the number of contact points varying in time, with a distribution of elastic forces between 
them. However, results of such a model are expected to depend on numerous parameters, which 
vary not only from one sample to another but also in time. In particular, the number of 
contact points and their elastic constants are hard to be controlled. 

The aim of this paper is to compare the instability of the uniform motion for one and two blocks. 
In this approach, the blocks are equivalent to the contact points. Our main goal is to prove that 
the instability of the uniform motion occurs for the same values of the model parameters for one 
and two blocks. When discussing this result, we are faced with almost all above mentioned 
difficulties in the interpretation. In reality, the case of two contact points is probably as rare 
as the case of one. However, it seems to us that at least the direction is proper. We hope that
our analytical and numerical results provide a basis for more extensive search. 

In subsequent section we show that for one block the analytical conditions for the Hopf bifurcation 
\cite{glen} are not fulfilled. The calculations are supplemented with some numerical results, 
reported in Section III. In Section IV we prove analytically, that the transition point is the same for the 
case of one and two blocks. Again, the simulations support these results. Last section is devoted 
to final conclusions.

\section{The equations of Rice and Ruina}

The equations of motion can be formulated \cite{bur,bauII} as follows

\begin{equation}
\frac{K}{W}(Vt-x)=\mu_0+B\ln \frac{\Phi}{\Phi_0}+A\ln \frac{\dot{x}}{V_0}
\end{equation}

\begin{equation}
\dot{\Phi}=1-\frac{\dot{x}\Phi}{D_0}
\end{equation}
where $x-Vt$ is the block position with respect to the driving mechanism, $V$ is the driving 
velocity, $K$ is the spring constant, $W$ is the normal force, $\mu _0$ is a reference value 
of the friction coefficient for steady sliding at some velocity $V_0$. During sliding, the 
microcontacts are refreshed, on average, after a distance $D_0$. The state of these microcontacts 
is described by the variable $\Phi $, which interpolates between the time of stick for the block
sticked and $D_0/V$ for steady sliding. Finally, $\Phi _0=D_0/V_0$ and $A$, $B$ are unitless material 
constants. We note that $B>A$ in the experimental data \cite{bau2}.

The equations can be transformed to an autonomous form. Denoting $(x-Vt)/D_0=\alpha$,
$\Phi /\Phi _0=\phi$, $V_0t/D_0=\tau$, $V/V_0=\omega$, exp$(-\mu _0/A)=\gamma$, $KD_0/W=\kappa$,
we get unitless equations

\begin{equation}
\dot{\alpha}=-\omega+\gamma\phi^{-\frac{B}{A}}\exp [-\frac{\kappa}{A}\alpha]
\end{equation}

\begin{equation}
\dot{\phi}=1-\gamma\phi^{1-\frac{B}{A}}\exp [-\frac{\kappa}{A}\alpha]
\end{equation}
where the time derivative is over $\tau$. These equations can be furter simplified by a change 
of variables: $-1/\omega$ exp$(-\kappa \alpha/A=x$, $\phi ^{-b-1}=y$, where $b=B/A-1$. Introducing
a parameter $\mu=b-\kappa/A$, and renormalizing time once more ($\tau \to \gamma \tau$)
we get the equations in more algebraic form

\begin{equation}\label{aa}
\dot{x}=(b-\mu)(mx-x^2y)
\end{equation}
\begin{equation}\label{ab}
\dot{y}=(1+b)(xy^2-\eta y^{1+\frac{1}{1+b}})
\end{equation}
where $m=\omega /\gamma >0$, $\eta=1/\gamma$. At the fixed point, where the block velocity is 
constant and equal to the driving velocity, $\dot{x}=0$ and $\dot{y}=0$. There,  

\begin{equation}
x=\eta(\frac{\eta}{m})^{b}
\end{equation}
\begin{equation}
y=\Big( \frac{m}{\eta} \Big) ^{(1+b)}
\end{equation}
The determinant of Jacobian $J$ at the fixed point is $(b-\mu)m^2$, and the trace is $\mu m$. This 
means that the stability of the fixed point is lost when the control parameter $\mu $ becomes 
positive \cite{glen}. Near this point, $Det(J)>0$.

In due course, only first derivatives over $\mu$ will be calculated at $\mu =0$. Then, 
writing down the eigenvalues of $J$ we can neglect terms proportional to $\mu ^2$.

\begin{equation}
\lambda=m(\frac{\mu}{2}\pm i\sqrt{b-\mu})
\end{equation}
The transformation to the Jordan form leads to new variables $\xi, \psi$ 

\begin{equation}
\xi=\frac{(1+b)}{m\sqrt{b}}(1+\frac{\mu}{2b})x+\sqrt{b}(\frac{m}{\eta})^{-2(1+b)}y
\end{equation}

\begin{equation}
\psi=(\frac{m}{\eta})^{-2(1+b)}y
\end{equation}
Now, the equations of motion are

{\setlength\arraycolsep{2pt}
\begin{eqnarray}
\dot{\xi}&=&\frac{m}{(1+b)}b\sqrt{b}(\frac{3}{2}\frac{\mu}{b}-1)(\frac{m}{\eta})^{2(1+b)}
(\xi-\sqrt{b}\psi)^2\psi-
mb(\frac{\mu}{b}-1)(\xi-\sqrt{b}\psi)+{}
\nonumber\\
& & {}+mb(1-\frac{\mu}{2b})(\frac{m}{\eta})^{2(1+b)}(\xi-\sqrt{b}\psi)\psi^2
-\eta \sqrt{b}(1+b)(\frac{m}{\eta})^2 \psi^{1+\frac{1}{1+b}}
\end{eqnarray}}

\begin{equation}
\dot{\psi}=m\sqrt{b}(1-\frac{\mu}{2b})(\frac{m}{\eta})^{2(1+b)}
(\xi-\sqrt{b}\psi)\psi^2
-(1+b)\eta(\frac{m}{\eta})^2\psi^{1+\frac{1}{1+b}}
\end{equation}

These equations can be written in short as $\dot{\xi}=f(\xi,\psi)$, $\dot{\psi}=g(\xi,\psi)$. 
The condition for the presence of the Hopf bifurcation \cite{glen} is 

\begin{eqnarray}
a&=&\frac{1}{16}(f_{\xi\xi\xi}+g_{\xi\xi\psi}+f_{\xi\psi\psi}+g_{\psi\psi\psi})+\nonumber \\
& &\frac{1}{16\omega}[f_{\xi\psi}(f_{\xi\xi}+f_{\psi\psi})-g_{\xi\psi}(g_{\xi\xi}+g_{\psi\psi})-
f_{\xi\xi}g_{\xi\xi}+f_{\psi\psi}g_{\psi\psi}]\ne 0
\end{eqnarray}
However, direct calculations for Eqns. (12,13) at the fixed point and $\mu=0$ lead to the 
result $a=0$.
 
As noted above, the phase of steady slip ceases to be stable when $\mu =B/A-1-KD_0/(WA)$ 
becomes positive. This can mean in particular, that the spring constant $K$ decreases. 
This is in accordance with the phase diagram in the creep regime, observed experimentally
\cite{bur,hes,bauII,ssc,nat}.

\section{Numerical calculations for one block}
\begin{figure}
\includegraphics[scale=0.92]{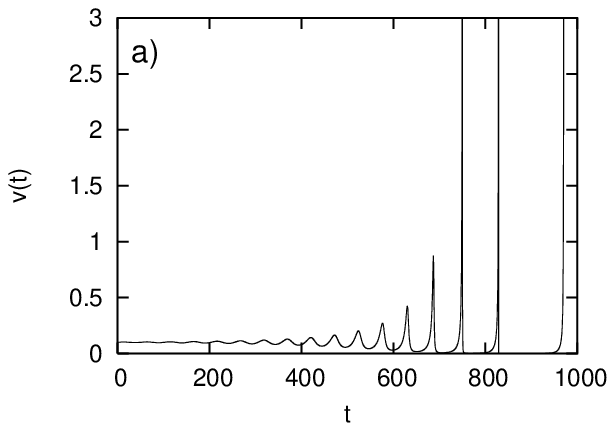}
\includegraphics[scale=0.92]{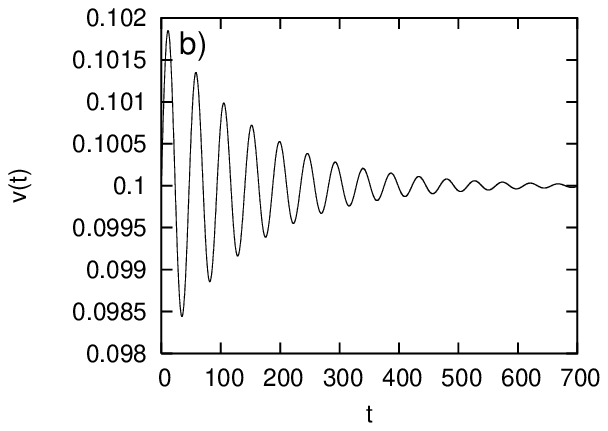}
\caption{Velocity of the single block versus time: (a) stick-slip (b)steady slip motion regime.}
\end{figure}

The stability of the solution of the equations of Rice and Ruina is checked numerically
at both sides of the transition. We applied the Runge-Kutta method of 4-th order. The
result is shown in Fig. 1. The parameters of the calculation \cite{book} are:
$B=0.08$, $A=0.03$,
$\mu_0=0.4$, $V=0.1 \mu m/s$,  $\phi_0=1.0$, $D_0=0.1 \mu m$, $K/W=0.054 \mu m^{-1}$ for the steady
slip and $K/W=0.046\mu m^{-1}$ for the stick-slip. The initial value of $\Phi$ 
is slightly
$(0.01)$ different from value characteristic for uniform movement with velocity equal to
driving mechanism velocity $V$. The initial values of block's position and velocity are 
equal to values characteristic for uniform movement.
As we see, either the steady slip or the stick-slip is observed, without an intermediate
phase with a continuous rise of the amplitude of the oscillations. Such a continuity is 
expected at the Hopf bifurcation.

\section{The case of two blocks}

In this case we rewrite the equations (1,2) twice with new variables 
$x_1,x_2,\Phi_1,\Phi_2$ and we add a coupling between the blocks by means of a new 
spring with constant $k_2$. Both driving springs have the same constants $k_1$. The 
equations are

\begin{equation}
\mu_0+B\ln \frac{\Phi_1}{\Phi_0}+A\ln \frac{\dot{x_1}}{V_0}=\frac{1}{W}(k_1(vt-x_1)+k_2(x_2-x_1))
\end{equation}

\begin{equation}
\mu_0+B\ln \frac{\Phi_2}{\Phi_0}+A\ln \frac{\dot{x_2}}{V_0}=\frac{1}{W}(k_1(vt-x_2)+k_2(x_1-x_2))
\end{equation}

\begin{equation}
\dot{\Phi_1}=1-\frac{\dot{x_1}\Phi_1}{D_0}
\end{equation}

\begin{equation}
\dot{\Phi_2}=1-\frac{\dot{x_2}\Phi_2}{D_0}
\end{equation}

Changes of variables, similar to the ones applied above, lead to autonomous dimensionless equations

\begin{equation}    
\dot{\alpha_1}=-\omega+\phi_1^{-\frac{B}{A}}\exp (-\mu_0) \exp (\frac{k_1D_0}{W}\alpha_1)  \exp (\frac{k_2D_0}{W}(\alpha_2-\alpha_1))
\end{equation}

\begin{equation}    
\dot{\alpha_2}=-\omega+\phi_2^{-\frac{B}{A}}\exp (-\mu_0) \exp (\frac{k_1D_0}{W}\alpha_2)  \exp (\frac{k_2D_0}{W}(\alpha_1-\alpha_2))
\end{equation}

\begin{equation}
\dot{\phi_1}=1-\gamma\phi_1^{1-\frac{B}{A}}\exp (-\mu_0) \exp (\frac{k_1D_0}{W}\alpha_1)  \exp (\frac{k_2D_0}{W}(\alpha_2-\alpha_1))
\end{equation}

\begin{equation}
\dot{\phi_2}=1-\gamma\phi_2^{1-\frac{B}{A}}\exp (-\mu_0) \exp (\frac{k_1D_0}{W}\alpha_2)  \exp (\frac{k_2D_0}{W}(\alpha_1-\alpha_2))
\end{equation}

The eigenvalues of the Jacobian at the fixed point are

\begin{equation}
\lambda_1=\frac{\omega}{2AW}[-D_0k_1-AW+BW-\sqrt{-4AD_0k_1W+(D_0k_1+AW-BW)^2}]
\end{equation}

\begin{equation}
\lambda_2=\frac{\omega}{2AW}[-D_0k_1-AW+BW+\sqrt{-4AD_0k_1W+(D_0k_1+AW-BW)^2}]
\end{equation}

{\setlength\arraycolsep{2pt}
\begin{eqnarray}
\lambda_3&=&\frac{\omega}{2AW}[-D_0k_1-2D_0k_2-AW+BW-\nonumber\\
& &\sqrt{(D_0k_1+2D_0k_2+AW-BW^2)-4AW(D_0k_1+2Dk_2)}]
\end{eqnarray}

{\setlength\arraycolsep{2pt}
\begin{eqnarray}
\lambda_4&=&\frac{\omega}{2AW}[-D_0k_1-2D_0k_2-AW+BW+\nonumber\\
& &\sqrt{(D_0k_1+2D_0k_2+AW-BW^2)-4AW(D_0k_1+2Dk_2)}]
\end{eqnarray}

Two of them, $\lambda_1$ and $\lambda_2$, change sign at the same value of $k_1/W$ where the 
transition for one block occurred. At this point,  $\lambda_3$ and $\lambda_4$ remain negative. 
This completes the proof that the transition point for two blocks coincides with the transition 
for one block. This is confirmed numerically, as shown in Fig. 2, for the same parameters as above, 
and $k_2/W=0.05 \mu m^{-1}$. In this case, the numerical plots reveal almost full synchronization 
of two blocks below and above the transition: the time dependences of the block velocities 
coincide.

\begin{figure}
\includegraphics[scale=0.92]{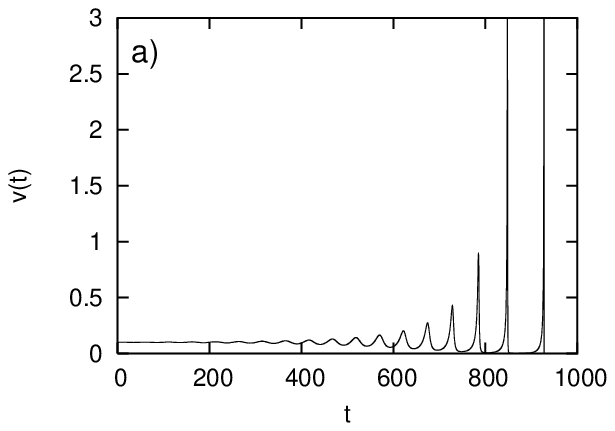}
\includegraphics[scale=0.92]{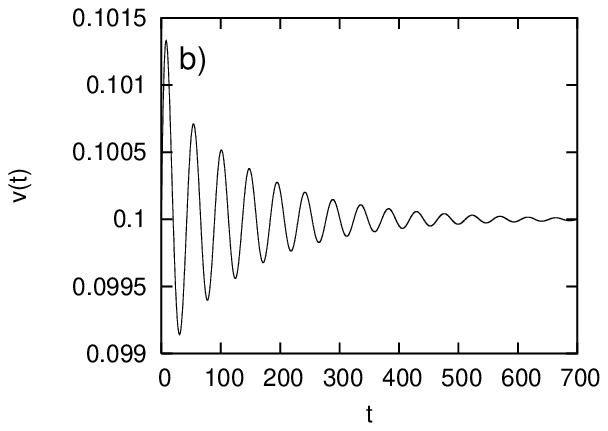}
\caption{Velocities of two blocks versus time: (a)slip-stick(b) steady slip motion regime.}
\end{figure}

\section{Conclusions}

We demonstrate that for two contact points, the uniform sliding ceases
its stability at the same value of the elastic driving force, that for one contact point.
This means, that the elastic force between the contact points, represented here as $k_2$,
does not influence the stability of the uniform motion. A question arises, if this result
could be valid for a larger number of blocks, i.e. of contact points. Such a generalization 
is of obvious interest; if true, it could release the limitation of the whole approach,
which in this case could apply to real surfaces with multiple contact points. However, 
we can only state that the question still remains open. 

For two blocks, our numerical results provide a demonstration of a synchronization of 
the stick-slip motion. However, we know that such a synchronization can depend on the 
initial conditions, and therefore it cannot be treated as generic cite{my}. This is true 
in particular if one tries to generalize the results for the larger number of contact points.

Acknowledgements. The authors are grateful to Tristan Baumberger for helpful comments.

\end{document}